\documentstyle[11pt,aaspp4]{article}
\tighten

\lefthead{XU, FANG \& DENG }
\righthead{Abundances and Correlations of Clusters}

\begin{document}

\title{Abundances and Correlations of Structures Beyond Galaxy Clusters}

\author{Wen Xu\footnote{
Beijing Astronomical Observatory, Beijing 100080,
China; also currently at Department of Physics and Astronomy, Arizona State
University, Tempe, AZ 85287},
Li-Zhi Fang\footnote{Department of Physics, University of Arizona,
Tucson, AZ 85721},
Zugan Deng\footnote{Department of Physics,
Graduate School, CAS, P.O. Box 3908,
Beijing, China}
}

\begin{abstract}

We investigated the structures on scales beyond the typical clusters of
galaxies. These structures are crucial to understand the cosmic
gravitational clustering in the pre-virialized stage, or quasilinear
r\`egime. Based on the multi-resolution analysis of the discrete wavelet 
transformation, we got statistical available ensembles of $r_{cl}$-clusters,
i.e. the structures on scale $r_{cl}$, in the range
$1 \leq r_{cl} \leq 24 \ h^{-1}$ Mpc for both N-body simulation
and the IRAS 1.2 Jy galaxy survey samples. If the mass-to-light ratio on 
scales larger than clusters asymptotically reaches a constant, we found that
the abundances and correlations of these IRAS $r_{cl}$-clusters 
to be basically consistent with the predictions of the flat low-density CDM
model (LCDM) and the open CDM model (OCDM), except the model-predicted
abundance of $r_{cl}$=24 $h^{-1}$ Mpc clusters seems to be higher than
IRAS data. The standard CDM (SCDM) gives too much power, and too weak
correlations on all the scales. For a given $r_{cl}$, the amplitude of
two-point correlation function of $r_{cl}$-clusters is increasing with
their richness. However, for a given richness (defined by the mean separation 
of neighbor objects), the clustering strengths of both simulation and
observation sample are found to be declining with $r_{cl}$ when $r_{cl}$ is
larger than 3 - 4 $h^{-1}$ Mpc. Therefore, the ``universal'' increase of 
the correlation amplitude with the scale of objects from galaxies, groups,
to poor, rich clusters is broken down for structures of
$r_{cl}>$ 3 - 4 $h^{-1}$ Mpc. Supercluster should not be a member of the 
``universal'' increase family.

\end{abstract}

\keywords{cosmology: theory -- galaxies: clusters: general
  --  large-scale structure of universe}

\section{Introduction}

The study of clusters of galaxies is among the most fruitful topics in 
cosmology. It provides rich information on large scale structures, significant 
for understanding the nature of formation and evolution of cosmic structures 
on scales of $\sim$ 1 - 10 Mpc. In particular, cosmological parameters 
such as the density parameter $\Omega_0$, the mass fluctuation amplitude
$\sigma_8$ and even the cosmological constant $\Omega_{\lambda}$ can be 
effectively constrained by observed properties of clusters. For instance, 
low mass density cold dark matter (CDM) models, $\Omega_0 \simeq 0 - 0.3$, 
are found to be generally favored by the abundance and two-point correlation 
function of clusters (Jing \& Fang 1994; Eke et al. 1996; Viana \& Liddle 
1996; Carlberg et al. 1997; Bahcall, Fan \& Cen 1997; Croft et al. 1997). 

However, the redshift and peculiar velocity data on scales larger than 
clusters, such as IRAS galaxy survey, yield $\Omega_0 \simeq 0.3 - 1.5$, with
a trend towards $\Omega_0 \simeq 1$ if moderate biasing is taken into account
(e.g. Dekel 1994; Strauss \& Willick 1995 for summary of the results).
Moreover, structures on scales larger than typical clusters are 
pre-virialized, or depend on the quasilinear evolution of cosmic 
clustering. Namely, the larger-than-cluster (LTC) objects
are in a quantitatively different dynamical evolution stage from the objects 
on scales equal to or less than Abell rich clusters which are virialized, 
or nearly virialized. Thus, one cannot estimate the clustering behavior 
of larger-than-cluster (LTC) objects by the results of clusters. Therefore, 
it is worth to investigate whether the low mass density CDM cosmogony is 
still favored by the abundance and two-point correlation function of 
structures on scales larger than clusters. This is the goal of this paper.

A curious thing is that we know of some observed LTC structures, like 
superclusters, and we know the dynamics of structure formation of popular 
models, but at the moment have not put them together. This is largely 
because the lack of a well-defined ensemble of LTCs. Structures on scales 
larger than clusters generally are not ``regular" in both configuration and 
velocity spaces. There is no well-defined dynamical criterion to classify 
and identify these structures, and therefore, statistical approaches 
have not yet been carefully pursued.

This problem can be solved by a {\it complete} decomposition. Since a
complete decomposition does not lose information, any ``irregular" objects
or distributions can be precisely represented by the coefficients of a 
complete decomposition. Thus, the model-observation confrontation can be 
done with a decomposed representation if the decomposition can reasonably 
work for both observed and model-predicted distributions. Recently it has 
been shown that a statistical analysis can indeed be performed for 
un-virialized large scale structures by a complete  multi-resolution
analysis of the discrete wavelet transformation ({\it DWT}). It has
successfully identified structures on scales from 1 to about 20 $h^{-1}$ 
Mpc, and revealed the scaling behavior of LTC clustering (Xu, Fang \& Wu
1998, hereafter Paper I).  

This paper is organized as follows. In \S 2 we present the abundances of 
IRAS $r_{cl}$-clusters and compare with the model predictions.
\S 3 tests the models with the IRAS two-point correlation functions of
the LTC structures. Finally, conclusion is in \S 4.

\section{Abundances of $r_{cl}$-clusters}

\subsection{Mass functions of $r_{cl}$-clusters from simulation samples}

The model samples of mass distributions used in this paper are generated by
N-body simulations with the P$^3$M code developed by Y.P. Jing
(Jing \& Fang 1994, Jing et al. 1995). Three popular models of the CDM 
cosmogony, i.e. the standard CDM (SCDM), the flat low-density CDM (LCDM) 
models and the open CDM model (OCDM), are employed. The Hubble constant 
$h$ ($\equiv H_0/(100$ km/sec/Mpc)), mass density $\Omega_0$, cosmological 
constant $\Omega_{\lambda}$ and the normalization of the power spectra of 
the density perturbation $\sigma_8$ are chosen to be (0.5, 1.0, 0.0, 0.62)
for SCDM, (0.75, 0.3, 0.7, 1.0) for LCDM, and (0.75, 0.3, 0.0, 1.0) for
OCDM. The simulation parameters are: 1.) box size $L=310\ h^{-1}$ Mpc; 2.)
number of simulation particles $N_p=64^3$ and 3.) the effective force
resolution $\eta=0.24\ h^{-1}$ Mpc. In this case, a particle has mass of
$3.14\times10^{13}\Omega_M h^{-1}M_{\odot}$, which is small enough to
resolve reliably objects with mass 
$M\geq 5.5 \times 10^{14} h^{-1} M_{\odot}$. We produced 8 realizations for 
each model. The mass of a $r_{cl}$-cluster is measured within a sphere of 
radius $r_{cl}$.

The method of finding clusters on scales $r_{cl}$ by discrete wavelet 
transformation (DWT) has been developed in Paper I (about the 
technical aspects of the DWT multi-scale identification, see also Fang 
\& Thews 1998). We identified clusters on scales 
$r_{cl} = 0.75, \ 1.5, \ 3, \ 6, \ 12,$  and $24 h^{-1}$Mpc from the 
simulated samples. A result is shown in
Fig. 1. It plots the top 25 massive $r_{cl}$-clusters for each scale of
$r_{cl}=0.75,\ 1.5,\ 3, \ 6,\ 12$, and 24 $h^{-1}$ Mpc, from one realization
of OCDM simulation with volume 310$^3 h^{-3}$ Mpc$^3$. 

Observationally, supercluster is defined as cluster of rich clusters (top 
$r_{ab}$-clusters). In Fig. 1, some $r_{ab}$-clusters are located together 
to form large scale clumps or filaments. These structures are obviously 
superclusters. Fig. 1 also shows, however, that some of the top 25 
$r_{cl}=24$ $h^{-1}$ Mpc halos contain nothing of 
$r_{cl} \leq $ 24 $h^{-1}$ Mpc massive halos. This is, although the mass 
and length scales of these $r_{cl}=24$ $h^{-1}$ Mpc clusters are the same 
as superclusters, they do not contain any of the top 25 clusters on smaller 
scales. This result indicates that the identification of superclusters by 
``cluster of rich clusters'' may miss some very massive objects  
on scale $r_{cl} =$ 24 $h^{-1}$ Mpc. 
Recently, it has been found that the 
dynamical properties of superclusters are not only determined by rich 
clusters, but also by numerous galaxies in the intra-cluster regions 
(Small et al. 1998). This also implies that one should directly identify 
LTC structures from matter or galaxy distributions, instead of using only 
rich clusters.

With the multi-scale identification of simulation samples, we have
model-predicted mass function of $r_{cl}$-clusters, $n(>M,r_{cl})$, which
is the number density of $r_{cl}$-clusters with mass larger than $M$.
The mass functions of clusters with $r_{cl}=1.5$, 3, 6 and 12 $h^{-1}$ Mpc
in models of the SCDM, LCDM and OCDM are shown in Fig. 2. The mass function 
of clusters on Abell scale, i.e. $r_{cl}=r_{ab}= 1.5$ $h^{-1}$ Mpc, is the 
same as that given by {\it friends-of-friends} identification 
(Jing \& Fang 1994). We have also
calculated mass functions of $r_{cl}=24$ $h^{-1}$ Mpc clusters, which
are not shown in Fig. 2, but will be employed below. The statistical errors 
of the  mass functions of $r_{cl}$-cluster are mainly caused by Poisson errors.
The middle part of the mass function is usually more reliable, as all the 
Poisson errors are relatively small in this range. In Fig. 2,
this middle range corresponds roughly to
$10^{-4} < n(>M, r_{cl}) < 10^{-7}$ $h^{3}$ Mpc$^{-3}$.

Since each cluster now is characterized by two parameters $M$ and 
$r_{cl}$, it is inconvenient to define richness by mass alone. Instead, the 
relative richness can be described by the mean neighbor separation $d$
between the clusters considered. The larger the $d$, the higher the 
richness.

\subsection{$r_{cl}$-clusters from  IRAS samples}

We have also applied {\it DWT} multi-resolution analysis on real samples of 
the redshift survey of the Infrared Astronomical Satellite (IRAS) galaxies 
with flux limit of $f_{60}\ge1.2Jy$ (Fisher et al. 1995). The IRAS 1.2 Jy 
sample contains 5313 IRAS galaxies, covering the whole sky with depth of about
$150 h^{-1}$ Mpc. 

The sampling rate of the IRAS sample is low. Therefore, the IRAS clusters
identified on Abell scale $r_{ab}$ contain much fewer galaxies than optical
clusters. This gives rise to higher Poisson errors. However, larger $r_{cl}$
clusters contain more members, and Poisson error is lower. The IRAS sample
can give statistically significant ensemble of $r_{cl}>r_{ab}$ clusters.
Moreover, the IRAS sample consists of less early type of galaxies, and more 
late type. Namely, it has less galaxies in the center of $r_{ab}$-clusters,
but more field galaxies. This is rather a good point for studying structures
beyond clusters, as late type galaxies probably are more fair tracers of mass 
field on larger scales.

To minimize the radial effects, we've divided the data into redshift intervals, 
or the velocity intervals of $[2500, 5000]$, $[5000, 7500]$, $[7500, 10000]$ 
and $[10000, 12500]$ km s$^{-1}$. We then identify $r_{cl}$-clusters in each 
interval. The details of cluster-finding algorithm has been discussed in
Paper I.

It has been shown that IRAS galaxies basically trace the local large scale 
structures. Some individual clusters and superclusters from optical surveys 
have also been identified from the IRAS data (Fisher et al. 1995; Webster et 
al. 1997). We checked this point further by cross-identification between the 
Abell clusters and the multi-scale identified $r_{cl}$ clusters of the IRAS
galaxies.
Using the position data ($l$, $b$,and $z$) collected in the compilation of
the Abell catalog (Abell, Corwin \& Olowin 1989), one can find 18 rich
clusters (with galaxy counts $>30$) in the range [5000, 7500] km s$^{-1}$.
They are A194, A347, A400, A426, A539, A569, A779, A1367, A1656, A2870,
A2877, A2911, A2933, A3389, A3537, A3656, A3706, A3879.

Since the maximum positioning uncertainty of our cluster finding algorithm is
about 3 $h^{-1}$ Mpc in 2-d, and 2000 km s$^{-1}$ in radial, we choose the
matching radius $r$ between the projections of centers of the
cross-identified clusters to be less than 3 $h^{-1}$ Mpc, and the radial
separation $d$ of the centers less than 2000 km s$^{-1}$. In this case,
7 out of the 18 Abell clusters are cross-identified with IRAS
$r_{cl}=3h^{-1}$Mpc clusters. These are A347, A426, A539, A1367, A3389,
A3656, A3706. The overall confidence level of this cross-identification
can be estimated by a binomial distribution function $B(k,n,p)$ that
a random distribution would generate $\geq k$ cross-identifications from
$n$ realizations when the chance alignment probability is known as $p$.
When overlapping is not important, $p =N\pi r^2d/V$, where $N$ and $V$ are
the total number of clusters and the volume of the redshift shell,
respectively.
For the above-mentioned Abell-IRAS cluster cross-identification the overall
confidence level is 3$\sigma$ ($\sim 99.7\%$).

The difference between 7 and 18 is due partially to the low sampling rate
of the IRAS, i.e. the higher Poisson errors. It is not a surprise that the
richness of individual clusters estimated from IRAS may have large
difference from optical estimate. This cross-identification seems to show
that the Abell clusters may be underestimated by the IRAS identification
by a factor about 2.6. However, we also did the cross-identification in the
opposite direction, i.e. to identify the 16 top IRAS clusters in the shell
[5000, 7500] km s$^{-1}$ with Abell clusters. The result is: 5 out of the 16
IRAS clusters have their Abell counterparts. Namely, the IRAS clusters may
be underestimated by the Abell clusters by a factor about 3. Therefore,
satistically, both Abell and IRAS clusters trace the same density field.
If we adjust the matching radius $r$ and separation $d$, we would find that
the number of Abell-IRAS cross-identified clusters increases from the 7 of 18
(3$\sigma$ confidence level), to 13 (2$\sigma$) and 15 (1$\sigma$).
This shows that the Abell and IRAS clusters provide about the same
description of the massive halos on scale $r_{ab}$.

Using the 7 better cross-identified $r_{cl}=3h^{-1}$Mpc IRAS-Abell clusters,
one can calculate the mean ratio of $N_{IRAS}/N_{abell}$, where
$N_{IRAS}=N_g(1.5)/\phi(z))$,
$N_{abell}$ is the number of galaxies of the considered Abell cluster,
$N_g(1.5)$ is the observed number of IRAS galaxies 
within radius 1.5 $h^{-1}$ Mpc
around the cluster center, and $\phi(z)$ the IRAS selection function
(Fisher et al. 1995). We found that the median value
$N_{IRAS}/N_{abell} = 7.03$. On the other hand, the mass-to-number ratio of
Abell clusters on average is
$M_{abell}/N_{abell} =0.6\times10^{13}h^{-1} M_{\odot}$,
where $M_{abell}$ is the mass of the Abell clusters
(Bahcall and Cen 1993). Thus, the total mass of the IRAS clusters $M$ can be
represented as
\begin{equation}
M= A N_{IRAS},
\end{equation}
where the number-mass conversion coefficient $A$ is 
\begin{equation}
A=10^{11.93} \   M_{\odot}.
\end{equation}

\subsection{Mass function of IRAS $r_{ab}$-clusters}

Considering the luminosity is proportional to the number of galaxies, 
the number-mass conversion relation eq.(1) essentially is the mass-to-light 
ratio for IRAS $r_{ab}$-clusters. Moreover, the mass-to-light ratio for 
Abell clusters is independent of the total mass of the clusters (Bahcall \& 
Cen 1993). Therefore, it seems to be reasonable to employ eq.(1) to find 
the total mass of IRAS $r_{ab}$-clusters. 

The total number of  member galaxies, $N_{IRAS}$, is given by 
$N_{IRAS} = N_g/\phi(z)$, where $\phi(z)$ is the selection function 
(Fisher et al. 1995). Thus, we have 
\begin{equation}
M = A N_g/\phi(z).
\end{equation}
Using eq.(3), the mass function of the IRAS $r_{ab}$-clusters can be found 
from their abundance $n_{nc}(>N_g,r_{cl})$ by
\begin{equation}
n_{IRAS}(>M, r_{ab})= n_{nc}(>N_g=M\phi(z)/A, r_{ab}).
\end{equation}

Using the $n_{nc}(>N_g, r_{ab})$ given by the IRAS galaxies in the shell
[5000, 7500] km s$^{-1}$, we calculated the mass function of the IRAS 
$r_{ab}$-clusters $n_{IRAS}(>M, r_{ab})$. We found that $n_{IRAS}(>M, r_{ab})$
is in very good agreement with the mass function of optical and X-ray 
clusters within Abell radius, $n_{op}(>M, r_{ab})$ (Bahcall and Cen 1992),
espicially for rich clusters of $M>10^{14}$ $h^{-1}$ $M_{\odot}$. 

We have also calculated the mass function $n_{IRAS}(>M, r_{ab})$ using
the IRAS data of the shells [2500, 5000], [7500, 10000] and 
[10000, 12500] km s$^{-1}$. The resulted $n_{IRAS}(>M, r_{ab})$ is also 
basically in agreement with $n_{op}(>M, r_{ab})$. The goodness of this 
agreement can be desrcibed by the number-mass conversion coefficient $A$. In 
order to have a best fitting of $n_{IRAS}(>M, r_{ab})$ with 
$n_{op}(>M, r_{ab})$, we find $\log A=12.07\pm0.21$ for shell [2500, 5000]
km s$^{-1}$, $11.93 \pm 0.15$ for [5000, 7500],  $11.71\pm0.32$ for 
[7500, 10000], $11.63\pm0.41$ for [10000, 12500]. Here the errors include 
both the Poisson errors of $n_{IRAS}$ and $n_{op}$. $A$ shows a slight 
decrease from lower velocity shell to higher velocity. This might be caused 
either by the uncertainty of the selection function, or by a very weak 
redshift variation of IRAS galaxies with respect to optical clusters. 

The conversion coefficients $A$ from all the redshift shells are the same as 
the $A$ of eq.(2) within errors, and their differences are not larger than a 
factor of 3. This is, in terms of mass function of $r_{ab}$-clusters, the 
matter distributions traced by the IRAS galaxies is about the same as optical 
and X-ray clusters on large scales. This result seems to be a little surprising.
It is known that the luminous IRAS galaxies with heavy dust lane do not 
appear at the densest regions, namely the clustering of infrared galaxies is 
weaker than that of optical galaxies at the core region of rich clusters. 
Therefore, the matter distributions represented by IRAS galaxies should be 
less clustering than that traced by optical galaxies on small scales. In 
other words, the mass-to-light ratio should actually be higher for IRAS 
clusters at the core region. However, the mass function given by eq.(4) 
matches well with that of the optical clusters. This is because the mass 
function eq.(4) measures statistically the mean number of collapsed massive 
$r_{ab}$-halos over a large volume. It is actually a measure of clustering 
features on scales equal to or larger than clusters. It thus less depends on 
the physical process within the core region. In a word, the IRAS galaxies may 
cause underestimates of individual masses of some richest clusters on smaller 
scales, while for large scales, IRAS galaxies give about the same 
statistically features of mass field as optical objects. 

\subsection{Model comparisons of abundances of IRAS $r_{cl}$-clusters}

Similar to eq.(4), one may derive the mass function of IRAS 
$r_{cl}$-clusters, $n_{IRAS}(>M,r_{cl})$, from their number-count
function $n_{nc}(>N_g, r_{cl})$ by
\begin{equation}
n_{IRAS}(>M, r_{cl})= n_{nc}(>N_g=M\phi(z)/A, r_{cl}).
\end{equation}
The problem is yet how to determine $A$ at scale $r_{cl}> r_{ab}$.

From eq.(1), $A$ essentially is the mass-to-light ratio of the system on
scale $r_{ab}$. The derived value of $A$ for $r_{ab}$-clusters corresponds 
to a median mass-to-light ratio of $M/L \simeq 300 \pm 100\ h (M/L)_{\odot}$.
As mentioned above, $r_{ab}$-clusters are dominated by elliptical galaxies, 
while the larger structures have higher fraction of spiral galaxies. The 
mass-to-light ratio or $A$ on scales $r_{cl}> r_{ab}$ may be smaller than 
$r_{ab}$. The observed Virgo cluster infall motion, however, indicates that 
$M/L$ ratio doesn't show a decrease with scale, while appears to flatten and 
remain approximately constant on scales larger than 0.2 $h^{-1}$ Mpc 
(Bahcall, Lubin \& Dorman 1995). This result is strengthened by a direct 
measure of  $M/L$ of the Corona Borealis supercluster  (Small et al. 1998). 
It gives $M/L \simeq 560\ h (M/L)_{\odot}$.  Namely, the $M/L$ ratio on scale 
$\sim$ 20 $h^{-1}$ Mpc is not lower, but even higher than that of rich 
clusters by a factor about 1.9. This implies that the $M/L$ ratio is weakly 
scale-dependent from scales of 1 to about 20 $h^{-1}$ Mpc. The weak dependence 
of $M/L$ on scale is also supported by the following fact: the segregation 
between elliptical and spiral galaxies of APM Bright Galaxies Catalogue (BGC), 
as well as the ratio between the second order bias factors of these APM-BGC 
elliptical and spiral galaxies, is found to be basically scale-independent 
from 1 to 20 $h^{-1}$ Mpc (Loveday et al. 1995, Fang, Deng \& Xia, 1998). 
It would be reasonable to assume that the elliptical-spiral mix doesn't cause a 
serious scale-dependence of the mass-to-light ratio on large scales, and 
$A$ is scale-independent within a factor of 3. 

Using the $A$ for each shell (\S 2.3), we calculated the mass functions of 
IRAS clusters with
$r_{cl}= 1.5, 3, 6, 12$ and $24 h^{-1}$ Mpc for data of the velocity
shells [2500, 5000], [5000, 7500], [7500, 10000] and
[10000, 12500] km s$^{-1}$. The errors of the IRAS
cluster mass functions are estimated by the same way as that of optical 
clusters. The vertical error bars in Figs. 3 - 6 are the 
Poisson errors of counting the clusters in the limited volume of a shell.
The horizontal error bars are the Poisson errors of mass estimation of each 
clusters. This Poisson error is generally smaller for larger $r_{cl}$, as 
more galaxies are contained in larger $r_{cl}$ clusters. Therefore, at larger 
$r_{cl}$ the IRAS data can set effective constraints.

Since the ratio between mass functions at $r_{cl}$ and  $r_{ab}$ is independent
of the value $A$, Figs. 3 - 6 essentially are tests of the $r_{cl}$-dependence
of mass functions of $r_{cl}$ clusters.  Despite the errors 
of these figures are large, Figs. 3 and 4 have shown that all $n(>M,r_{cl})$ 
in the range of $r_{cl} \leq 12$ $h^{-1}$ Mpc and in the two shells 
[2500, 5000] and [5000, 7500] km s$^{-1}$ are in good agreement with the 
predictions of the models of LCDM and OCDM. The results of 24 h$^{-1}$ Mpc 
in these two shells (which have not directly been shown in Figs 3 and 4) 
are also basically consistent with LCDM and OCDM. The dynamical status of 
$r_{ab}$-clusters is very different from $r_{cl}>>r_{ab}$ clusters. The 
former are most likely virialized, the later are not. Therefore, Figs. 3 and 4
indicate that the models of OCDM and LCDM are still in good shape for 
clustering in pre-virialization regimes.

For shells of [7500, 10000] and [10000, 12500] km s$^{-1}$, Figs. 5 and 6
show that the LCDM and OCDM can fit with the data of $r_{cl} \leq 6$ $h^{-1}$ Mpc. 
Yet, the IRAS mass functions on scales of $r_{cl} = 12$ and 24 $h^{-1}$ Mpc 
are lower than the predictions of LCDM and OCDM. The deficit of IRAS 
$r_{cl} = 24$ $h^{-1}$ Mpc clusters may result from the poorness of the current data 
at $> 7500$ km s$^{-1}$. Since the overdensity of the large scale objects 
is relatively low, their detection will be poorer than small scales. 
Moreover, not all predicted $r_{cl}= 12,$ 24 $h^{-1}$ Mpc clusters contain 
top rich $r_{ab}$-clusters (Fig.1), and therefore, some of these structures 
will be overlooked if the completeness of samples is not high enough.   

Figs. 3-6 clearly show a systematic and significant lower of the IRAS mass 
functions in all velocity shells than the predictions of SCDM. SCDM gives 
too many massive clusters on ${\it all}$ scales of 
$r_{cl} \leq 24$ $h^{-1}$ Mpc. It is interesting to note that if all 
IRAS mass functions are shifted along the $\log M$ axis by a factor 
of about $0.60$, the SCDM model will be able to survive. The shifting factor 
0.60 requires $A$ to be larger than our derived value by a factor 
$10^{0.6}\simeq 4$, and the corresponding ratio $M/L$ should be as high as 
$4\times 300 \simeq 1200 h (M/L)_{\odot}$.
Thus, the SCDM can not be saved if the 
uncertainty of $A$ is less than a factor of 4.

\section{Two Point Correlation Functions of $r_{cl}$-Clusters}

\subsection{$d$- dependence of correlations}

It has been known for a long time that the two-point cluster correlation 
function $\xi_{cc}(r)$ follows the same power law as galaxies, 
i.e. $\xi_{cc}(r)=(r/r_0)^{-\gamma}$ and $\gamma \sim 1.8$, but has much
larger correlation length $r_0$ than galaxies. Moreover, the correlation
length $r_0$ defined by $\xi(r_0) \equiv 1$ is found to be systematically 
increasing with the cluster richness. An empirically ``universal'' scaling  
$r_0=0.4 d$, $d$ being the mean neighbor separation as employed to describe
the cluster richness (\S 2.1), was proposed to fit with the correlations 
of galaxies, QSOs, group of galaxies, X-ray and Abell clusters and even 
superclusters (e.g. Bahcall \& Soneira 1983; Bahcall \& West 1992 and 
references therein). Because the sizes of these objects, from galaxies to 
superclusters, are systematically increasing by at least two orders, it is 
unclear whether various objects with very different scales can really be 
fitted with one ``universal'' scaling. 

This problem can be solved by studying the two-point correlations of
clusters with different $r_{cl}$. Using samples of $r_{cl}$-clusters
from N-body simulation, we found that the two-point correlation functions
of $r_{cl}$-clusters can be well fitted with a power law
$\xi(r)=(r/r_0)^{-\gamma}$ for all considered $r_{cl}$ and $d$. The index
$\gamma$ is in the range of 1.8 - 2.0 for the LCDM  and OCDM, and
1.9-2.1 for the SCDM. No indication shows that the index $\gamma$ would 
significantly vary with either $r_{cl}$ or $d$.

The $r_0$-$d$ relations for $r_{cl}$-clusters are calculated. The results 
of the OCDM are plotted in Fig. 7. This figure shows that the $r_0$-$d$ lines 
for $r_{cl}= 0.75$, 1.5 and 3 $h^{-1}$ Mpc almost coincide. This is, in the
range of $r_{cl} \leq $ 3 $h^{-1}$ Mpc, the $r_0$-$d$ relation is 
``universal'', independent of the size of structures. The model SCDM and 
LCDM show similar ``universal'' $r_0$-$d$ relations.

However, the $r_0$-$d$ relation is substantially dependent on $r_{cl}$ if 
$r_{cl} >$ 3 $h^{-1}$ Mpc. Despite the correlation length $r_0$ is still 
increasing with richness $d$, the lines $r_0-d$ are lower for larger 
$r_{cl}$. Therefore, superclusters are not a member of the universal 
$r_0$-$d$ relation of $r_{cl} \leq $ 3  $h^{-1}$ Mpc clusters.

For clusters of $r_{cl}$ larger than 12 $h^{-1}$ Mpc, their
$r_0$ is comparable to and even less than $r_{cl}$. In this case, $r_0$ is
actually determined by an extrapolation of the two-point correlation
function $\xi(r)$ from range $r> r_{cl}$ to $r<r_{cl}$, and solving
$\xi(r_0)=1$. This method might be susceptible to large error. 
Nevertheless, it is reliable that lines of $r_0-d$ are significantly lower 
than the $r_0-d$ of smaller $r_{cl}$ clusters.

In redshift space the correlation length will increase by no more than 
$2 h^{-1}$ Mpc, and the index $\gamma$ becomes slightly larger. Our overall 
conclusions on correlation functions will not change due to these slight
differences.

\subsection{$r_{cl}$-dependence of correlations}

Fig.8 plots the correlation length $r_0$ as a function of $r_{cl}$ for a 
given $d$. This figure, actually, is an alternative representation of Fig.7.
It shows that for a given $d$ the correlation length $r_{0}$ is
almost invariant for $r_{cl}$ equal to or less than 3 $ h^{-1}$ Mpc,
but slowly decreasing with $r_{cl}$ when $r_{cl}$ is larger than about 4
$h^{-1}$ Mpc. This is a common feature for all the three models.

Theoretically, the decrease of $r_0$ with $r_{cl}$ is expected. Fig. 1 shows
that most halos of $r_{cl}=0.75$, 1.5 or 3 $h^{-1}$ Mpc coincide to form
a 0.75-1.5-3 clouds-in-clouds. This is because most halos of
$r_{cl} \leq 2 $ $h^{-1}$ Mpc are virialized. Therefore, the detection
of $r_{cl}$-clusters with $r_{cl}\leq 3$ $h^{-1}$ Mpc will find almost 
the same objects. In other words, the ensembles of 
$r_{cl}=0.75$, 1.5 or 3 $h^{-1}$ Mpc clusters actually consist of the same 
objects, and therefore, they have the same correlations, and the universal 
$r_{0}$-$d$ relations. However, on scales larger than 4 $h^{-1}$ Mpc, more 
and more halos are less developed, and their correlations are lower. 
Therefore, the ``universal'' increase of the correlation amplitudes with 
scale from galaxies, groups to clusters should be broken down for 
$r_{cl}>$ 3 - 4 $h^{-1}$ Mpc clusters. Superclusters are not members of 
the ``universal'' scaling family.

\subsection{Correlation lengths of IRAS clusters}

With the samples of the IRAS $r_{cl}$-clusters, we calculated the correlation
length $r_0$ by fitting the observed data with $\xi(r)=(r/r_0)^{-1.8}$.
The correlation lengths $r_0$ are measured from data of different velocity 
shells. The results of $r_0$ vs. $d$ are plotted in Fig. 9. The data points 
shown in Fig. 9 are binned in $d$. The errors are given by the variance of 
the correlation strengths at each bin of $d$. For $r_{cl}=r_{ab}$, 
the values of $r_0$ found from the IRAS samples are in agreement
with those from APM sample (Bahcall \& Cen 1992; Croft et al. 1997). 
Fig. 9 shows, obviously, that the $r_0$-$d$ relation from IRAS samples of 
$r_{cl}=$ 1.5 to 12 $h^{-1}$ Mpc favors the models of OCDM and LCDM. The 
SCDM-predicted $r_0$-$d$ curves are systematically lower than the IRAS results 
for all considered $d$.

We also tested the models by the $r_{cl}$-dependence of $r_0$. Fig.10 
gives the $r_0$-$r_{cl}$ relations from the simulation samples and IRAS data. 
The errors of the IRAS data are also large. But the observed data show
once again that the LCDM and OCDM are basically consistent with the IRAS
data, while the SCDM-predicted $r_0$ is too low to match with the observation.
More importantly, it is clear to see the slow decrease of $r_{0}$ with the
increase of $r_{cl}$ in all models. The $r_0$ decrease is an indicator of 
the transform from quasilinear to non-linear r\'egimes.

\section{Conclusions and discussions}

Structures on scales equal to and larger than typical clusters have been
identified by the {\it DWT} decomposition. This method is shown to be able to
systematically and uniformly identify structures on various scales, especially
those of larger than the size of typical clusters. Taking this advantage,
one can produce statistically available ensembles of $r_{cl}$-clusters from
simulation samples and observational data. 

To describe the dynamics of the cosmic clustering on scales beyond clusters, 
we introduced several new statistical measures: the bivariant abundance
function $n(>M,r_{cl})$, the $r_{cl}$-dependence of abundance, the
$r_{cl}$-dependence of correlations, etc. These statistics are useful to
reveal the features of pre-virialized regime of cosmic gravitational 
clustering, and to discriminate among models of structure formation, 
especially on large scales.

The identification of LTCs is effective to extract information from sparsely 
distributed data like the IRAS galaxies. The IRAS 1.2 Jy galaxies have low 
sampling rate compared to optical galaxies and are not found at cluster 
centers. They are not very effective to be employed for identification of 
traditional $r_{ab}$-clusters. However, the IRAS clusters with 
$r_{cl} > 3$ $h^{-1}$ Mpc can already set effective constraints on models, if 
the mass-to-light ratio for structures traced by IRAS galaxies is assumed to 
asymptotically reach a constant on scales larger than Abell radius. 
Therefore, IRAS galaxies can provide the same information of large scale 
structures as optical and X-ray galaxies, if the considered scales are larger 
than the scale of morphological segregation of IRAS galaxies that no dusty 
spirals like IRAS sources exist at the immediate neigbhourhood of densest 
regions.

The multi-resolution  analysis of the IRAS sample 
concludes that the abundances $n(>M,r_{cl})$ and correlations of objects on 
scales $r_{cl}=$ 1 to 24 $h^{-1}$ Mpc can be described by LCDM and OCDM, while 
the SCDM is in difficulty. Namely, the abundances and two-point correlation 
functions of structures on scales as large as about 20 $h^{-1}$ Mpc favor
a low-density universe if $M/L$ ratio is assumed to be flat over these scales.

\acknowledgments

We would like to thank an anonymous referee for a detailed report
which improves the presentation of the paper. Wen Xu thanks the World 
Laboratory for a scholarship. This project was done when WX was visiting at 
Physics Department, University of Arizona.


\figcaption{A projected distribution of $r_{cl}$-clusters identified from
a realization of OCDM N-body simulation within  box 310 $h^{-3}$ Mpc$^3$.
It shows the top 25 of the richest clusters for each $r_{cl}$. The legends 
of the symbols are: plus for 0.75,  diamond for 1.5,  circle for 3, 
triangle for 6, square for 12 and giant circle for 24 $h^{-1}$ Mpc.
The linear sizes of symbols are roughly equal to their $r_{cl}$, but
slightly amplified for small $r_{cl}$ for clarity.
\label{fig1}}

\figcaption{Mass functions of clusters identified with radii
$r_{cl}=$ 1.5, 3.0, 6.0 and 12 $h^{-1}$ Mpc for SCDM, LCDM and OCDM models 
at redshift $z=0$. $n(>M,r_{cl})$ is the number density of $r_{cl}$-clusters 
with mass larger than $M$. $M$ is in unit of $h^{-1}$ M$_{\odot}$.
\label{fig2}}

\figcaption{Mass functions of IRAS $r_{cl}$-clusters. $n$ is in unit
of $h^{3}$ Mpc$^{-3}$. The horizontal errors are Poisson errors from
the number counting of galaxies in a given cluster, and the vertical 
errors are the Poisson errors from counting the clusters. The model 
predictions of SCDM, LCDM, OCDM are shown as dashed lines, thin solid
lines, and thick solid lines, respectively. Observed data are from the 
IRAS sample in the velocity shell [2500, 5000] km s$^{-1}$.
\label{fig3}}

\figcaption{The same as Fig. 3, but the data are from shell 
[5000, 7500] km s$^{-1}$. 
\label{fig4}}

\figcaption{The same as Fig. 3, but the data are from shell 
[7500, 10000] km s$^{-1}$. 
\label{fig5}}

\figcaption{The same as Fig. 3, but the data are from shell 
[10000, 12500] km s$^{-1}$. 
\label{fig6}}

\figcaption{The $r_0$-$d$ relations for clusters with different
$r_{cl}$ in the OCDM model.  The error bars of the curves of
$r_{cl} =$ 6 and 12 $h^{-1}$ Mpc are from the scattering among 
realizations of simulation.
\label{fig7}}

\figcaption{The $r_0$-$r_{cl}$ relations at different richness
for models of OCDM, LCDM, SCDM.
\label{fig8}}

\figcaption{The $r_0$-$d$ relations for $r_{cl}=$
1.5, 3, 6, 12$h^{-1}$ Mpc. The dashed, thin solid, and thick solid lines
are, respectively, the simulation results of models SCDM, LCDM and OCDM. 
The observed data are from the averages over the velocity shells of the
IRAS galaxies. 
\label{fig9}}

\figcaption{The $r_0$-$r_{cl}$ relations for $d=$
25, 50, 70 $h^{-1}$ Mpc. The dashed, thin solid, and thick solid lines
are, respectively, the simulation results of models SCDM, LCDM and OCDM. 
The observed data are from the averages over the velocity shells of the 
IRAS galaxies. 
\label{fig10}}


\clearpage
\begin{thebibliography}{}
\bibitem{ab} Abell, G. O., Corwin, Jr. H. G. \& Olowin, R. P. 1989, \apjs, 70, 1.

\bibitem{18} Bahcall, N. A. \& Cen, R. Y. 1993, \apjl, 407, L49.

\bibitem{5} Bahcall, N. A., Fan, X. \& Cen, R. Y. 1997, \apjl, 485, L53.

\bibitem{ba} Bahcall, N. A., Lubin, L. M. \& Dorman, V. 1995, \apjl, 447, L81.

\bibitem{bs} Bahcall, N. A.,  \& Soneira, R. M. 1983, \apj, 270, 20.

\bibitem{ca} Carlberg, R. G., Morris, S. M., Yee, H. K. C. \&
         Ellingson, E. 1997, \apjl, 479, L19.

\bibitem{19} Croft, R. A. C., Dalton, G. B., Efstathiou, G., 
	Sutherland, W. J. \& Maddox, S. J. 1997, \mnras, 291, 305.

\bibitem{de} Dekel, A., 1994, \araa, 32, 371.

\bibitem{8}Eke, V. R., Cole, S., Frenk, C. S. \& Navarro, J. F. 1996,
	\mnras, 281, 703.

\bibitem{fa} Fang, L. Z., Deng, Z. G. \& Xia, X. Y. 1998, \apj, 506, 53.

\bibitem{fang} Fang, L. Z. \& Thews R. 1998, {\it Wavelts in Physics},
World Scientific, Singapore.

\bibitem{fi} Fisher, K. B. et al. 1995, \apjs, 100, 69.

\bibitem{2}Jing, Y. P. \& Fang, L. Z. 1994, \apj, 432, 438.

\bibitem{jing} Jing, Y. P., Mo, H. J., B\"orner, G. \& Fang, L. Z. 1995, \mnras, 274
, 417.


\bibitem{lo} Loveday, L., Maddox, S. J., Efstathiou, G. \& Peterson, B. A.
             1995, \apj, 442, 457.

\bibitem{ma} Small, T., Ma, C.-P., Sargent, W., Hamilton, D. 1998, \apj, 492,
45

\bibitem{sw} Strauss, M.A. \& Willick, J.A. 1995, Phys. Rep., 261, 271

\bibitem{vi} Viana, P. P. \& Liddle, A. R., 1996, \mnras, 281, 323.

\bibitem{we} Webster, M., Lahav, O., \& Fisher, K. 1997, \mnras, 287,
425.

\bibitem{xu} Xu, W., Fang, L. Z. \& Wu, X. P. 1998, \apj, 508, 472. (Paper I)

\end{thebibliography}
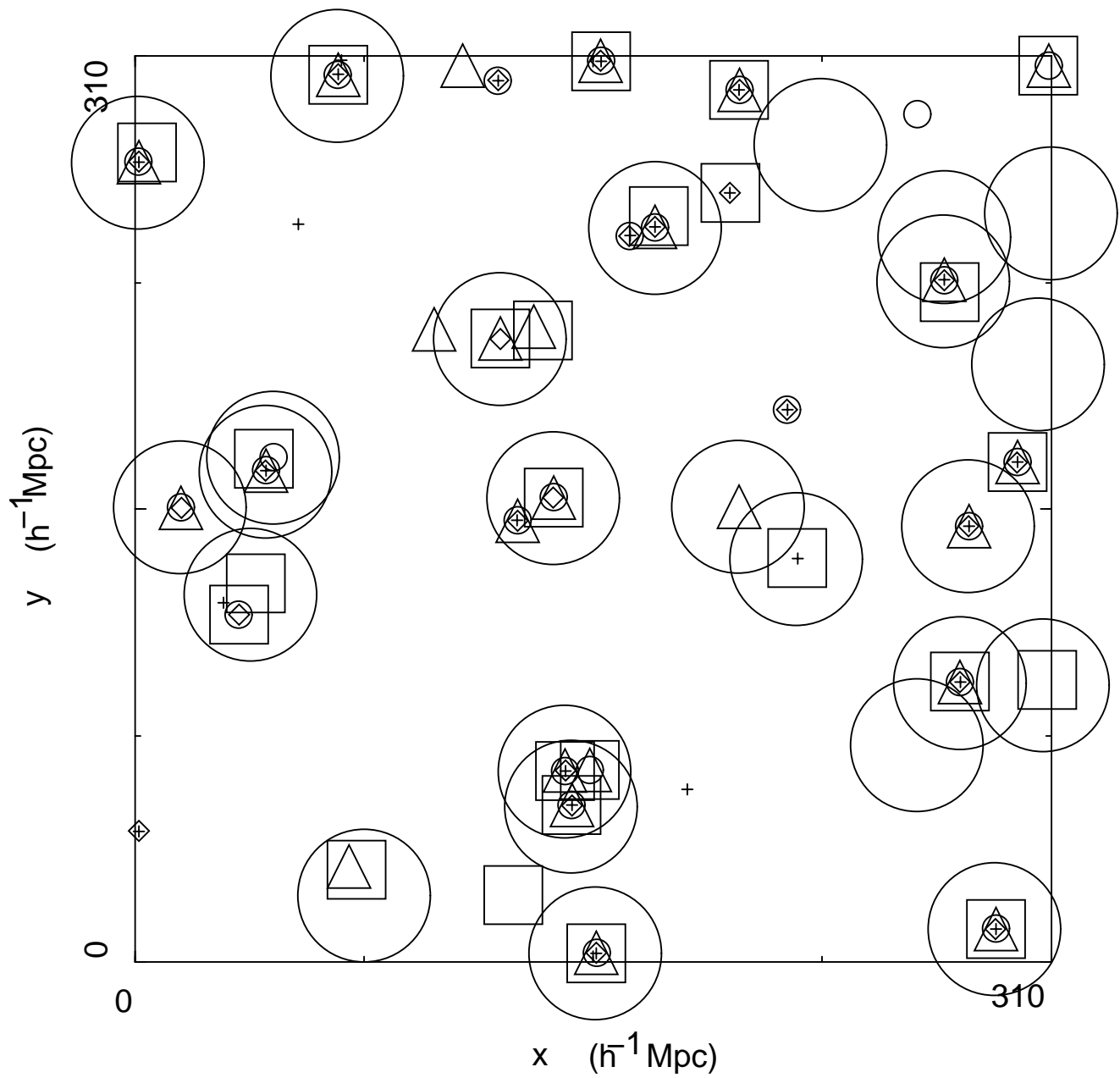
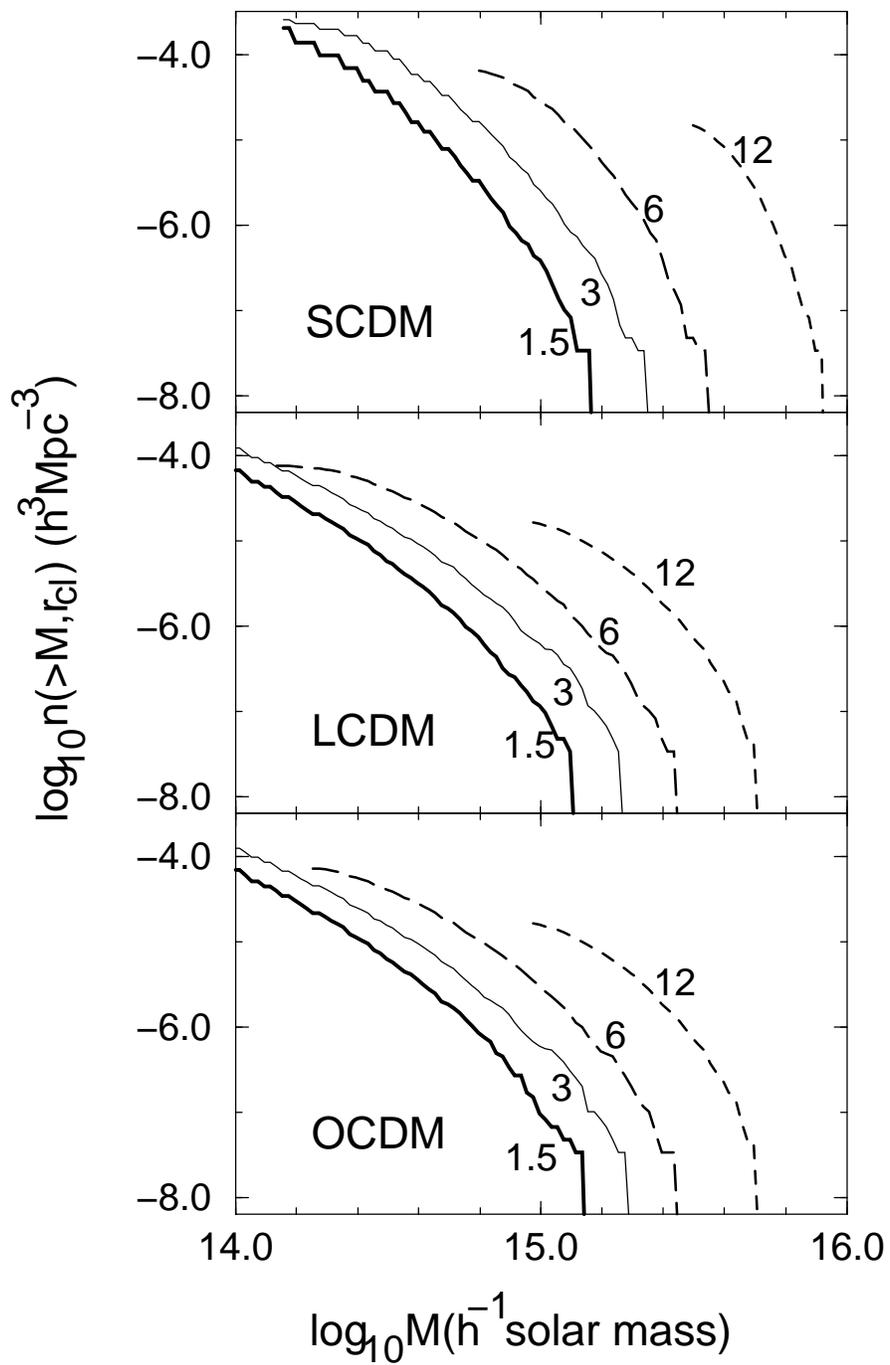
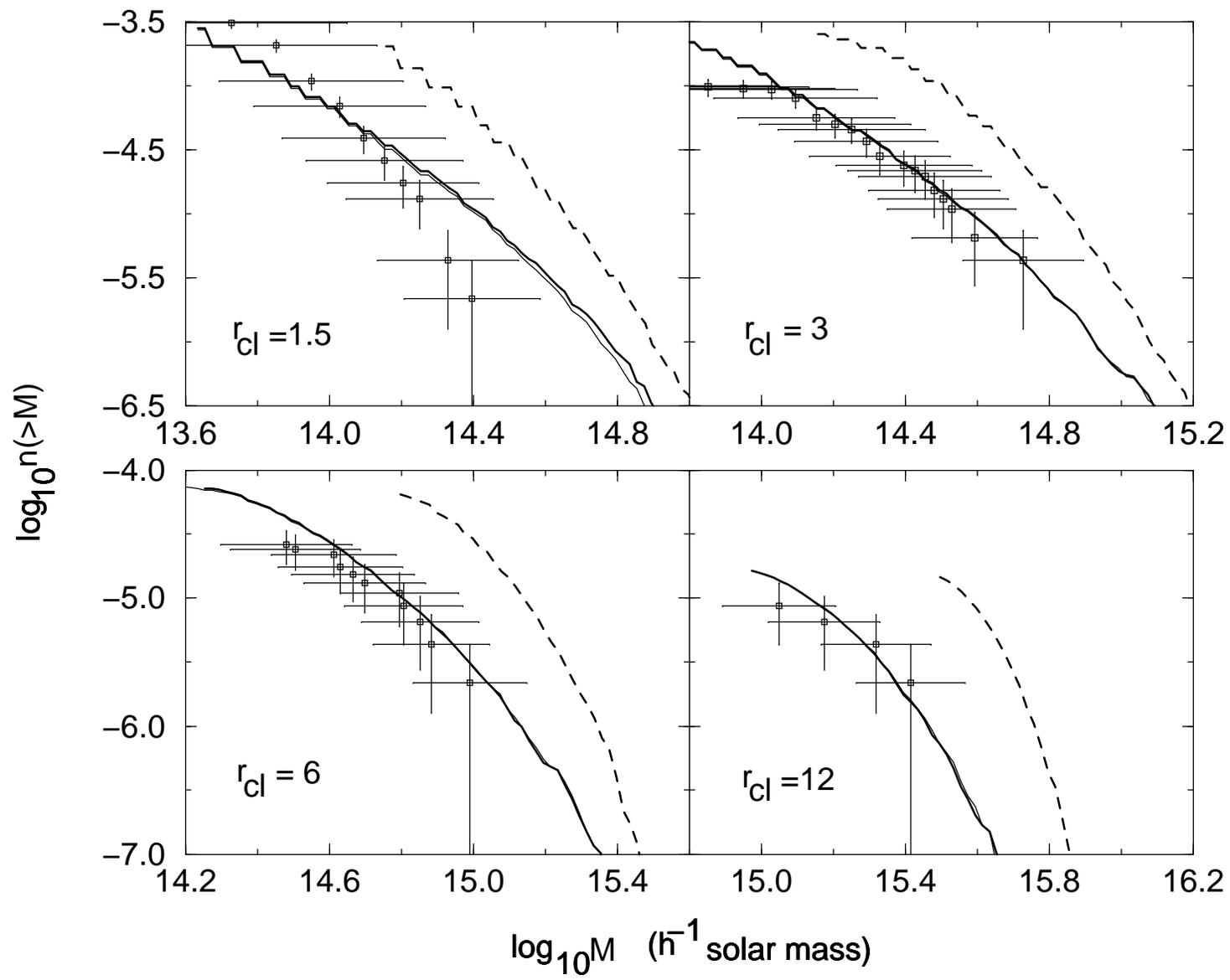
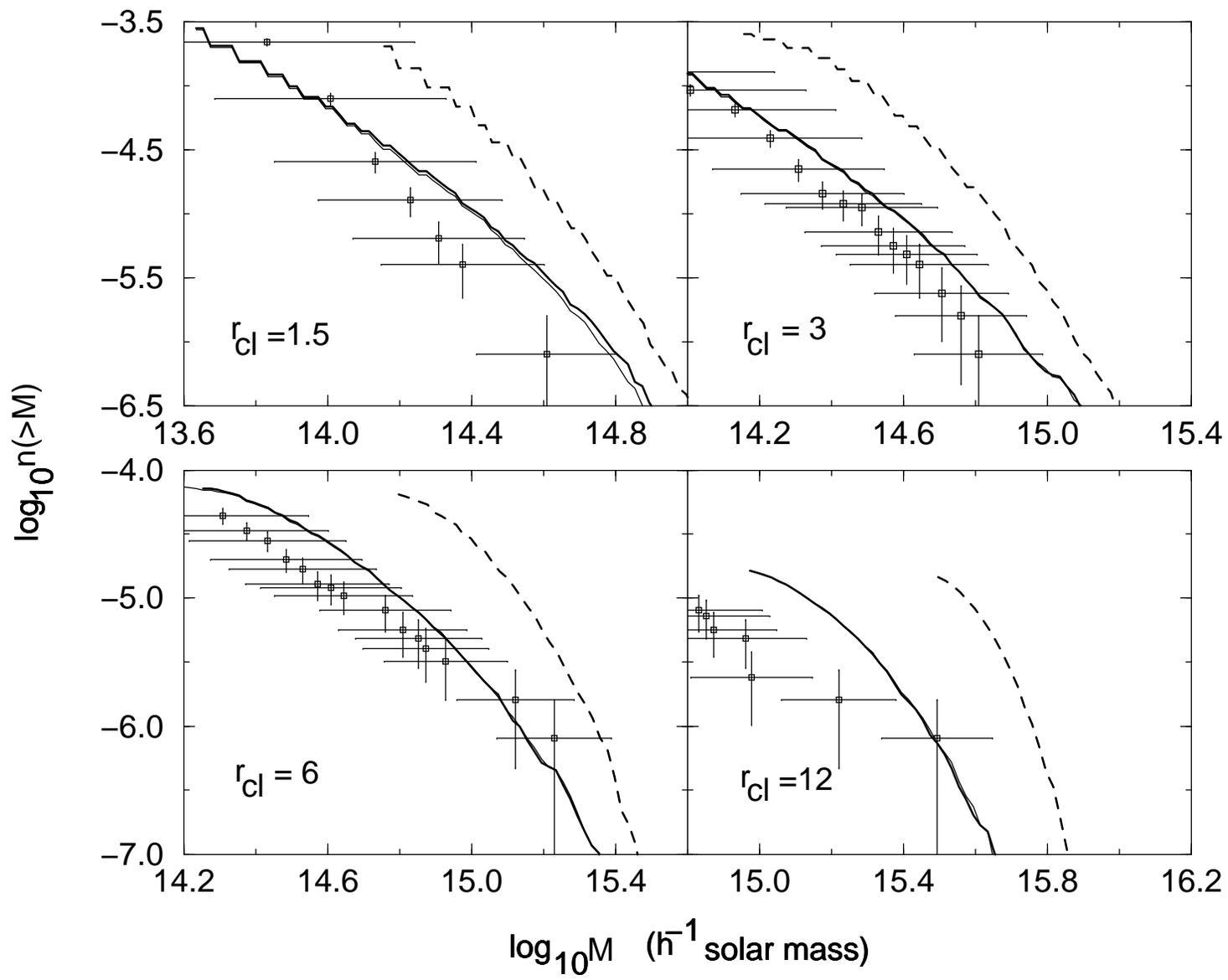
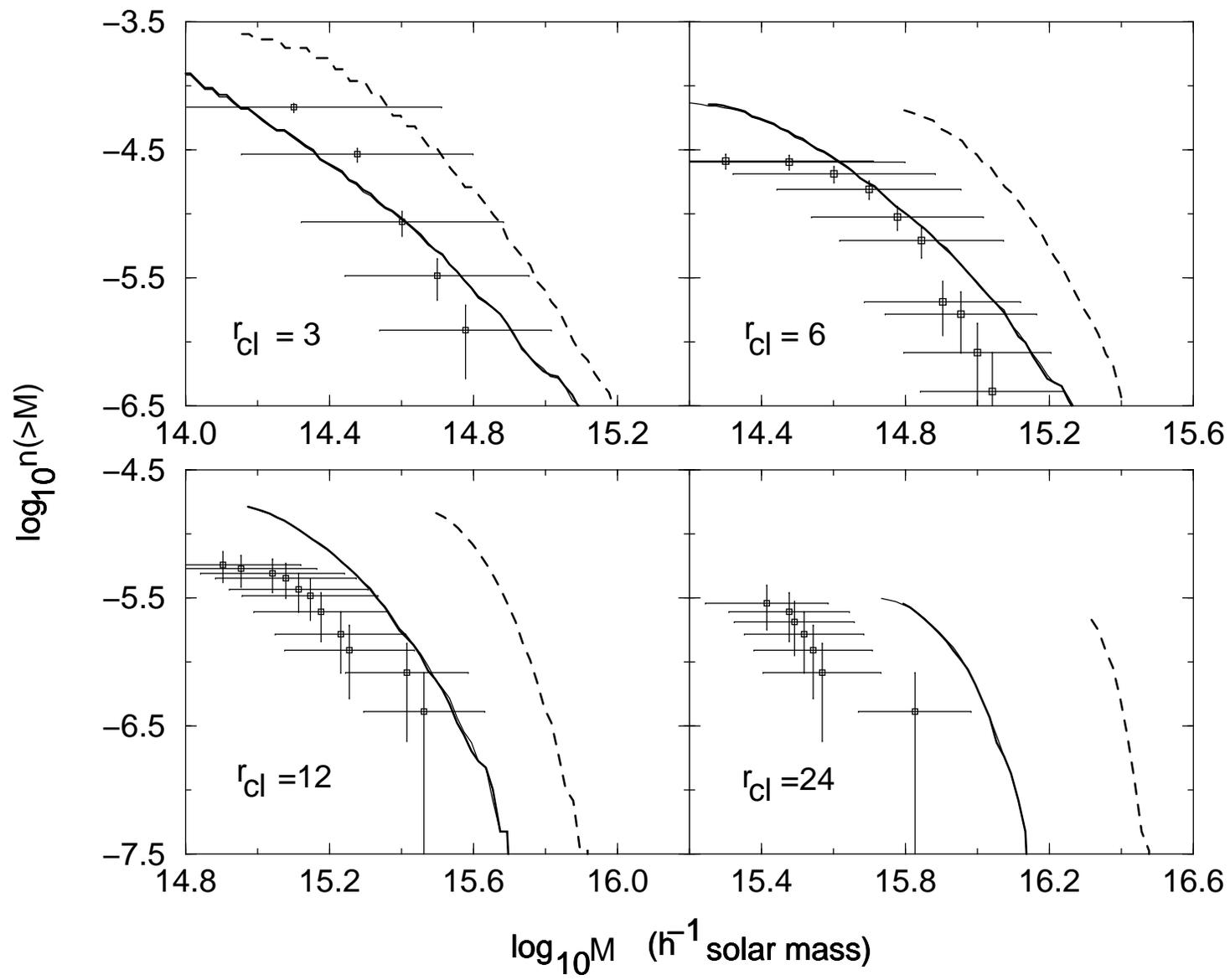
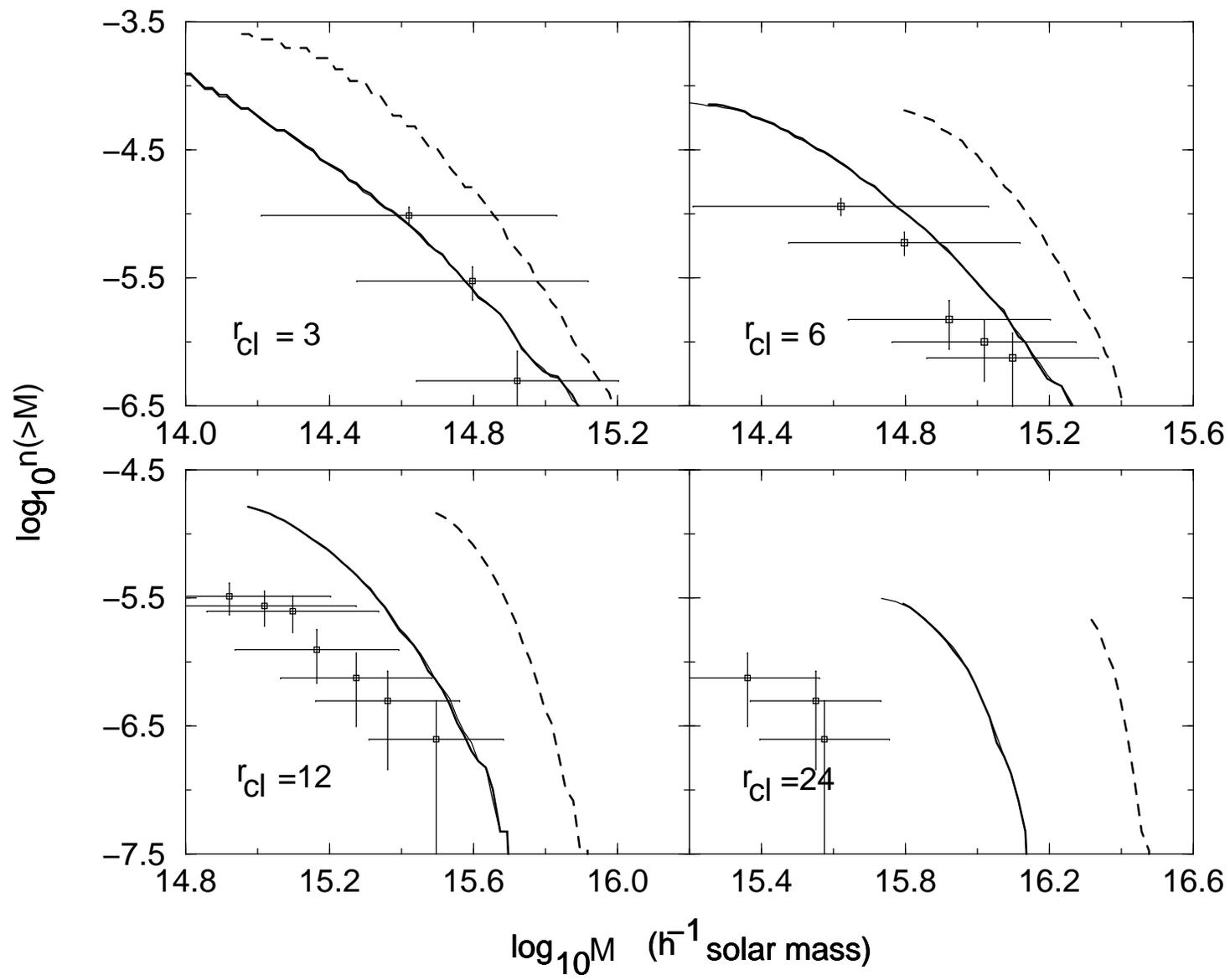
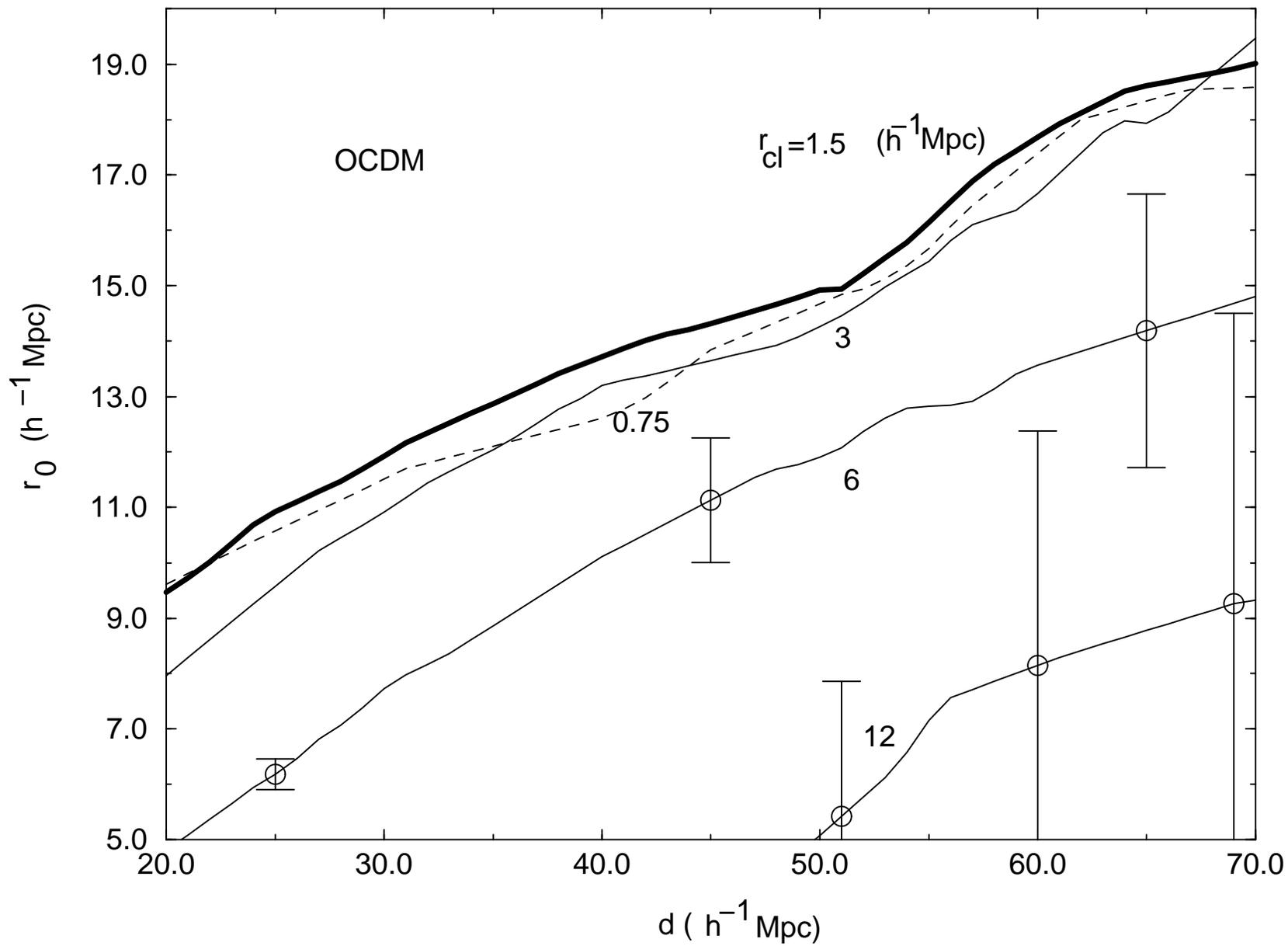
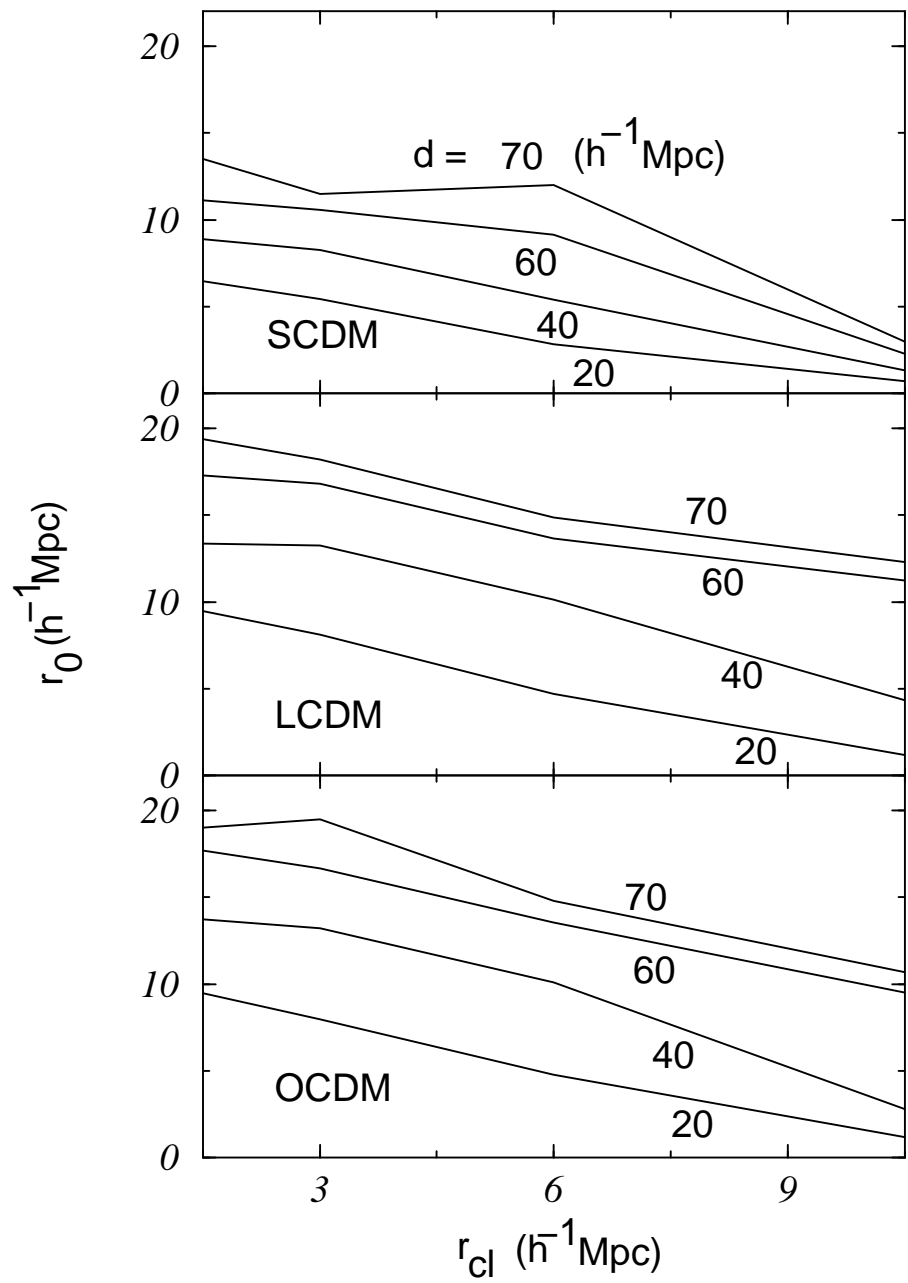
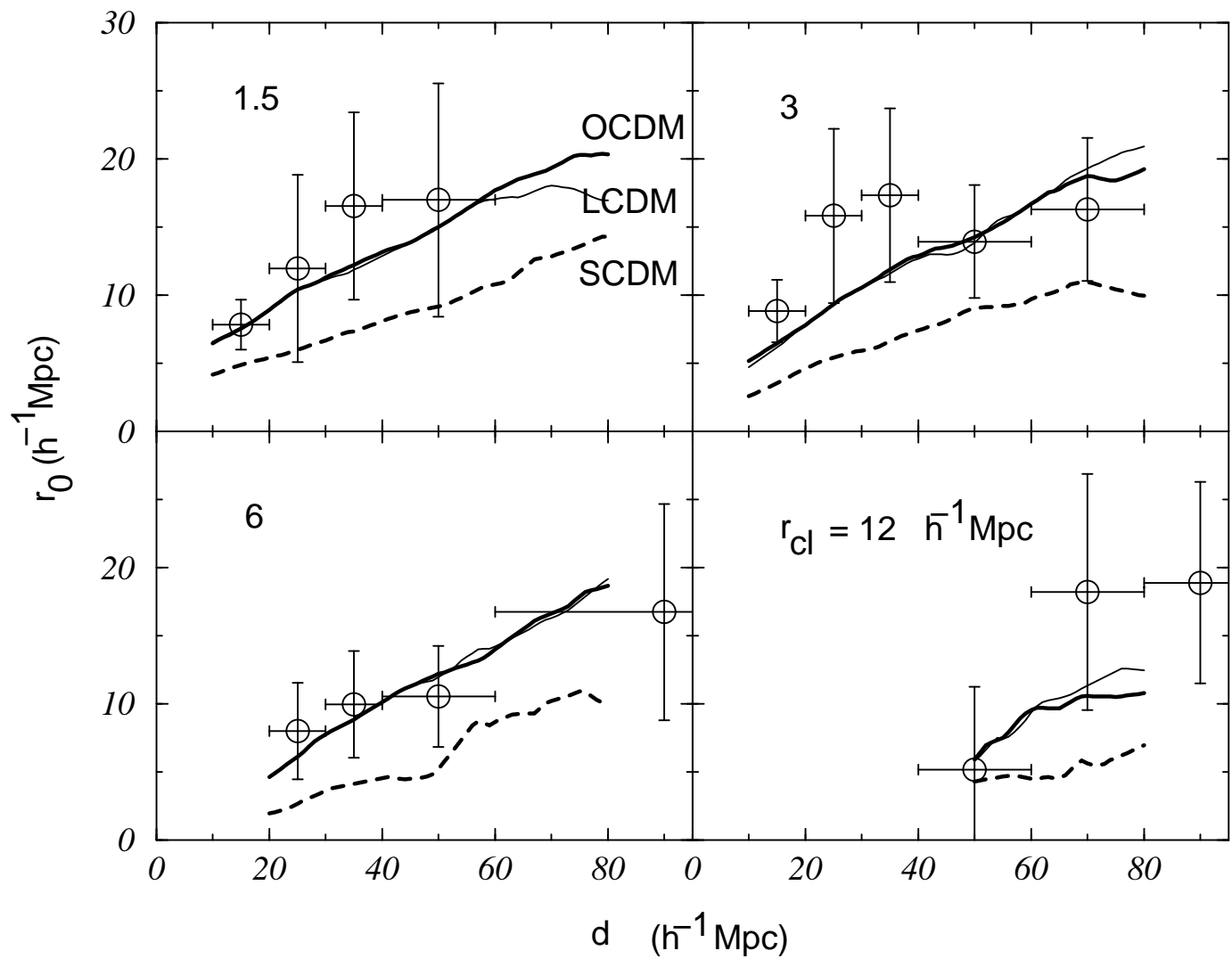
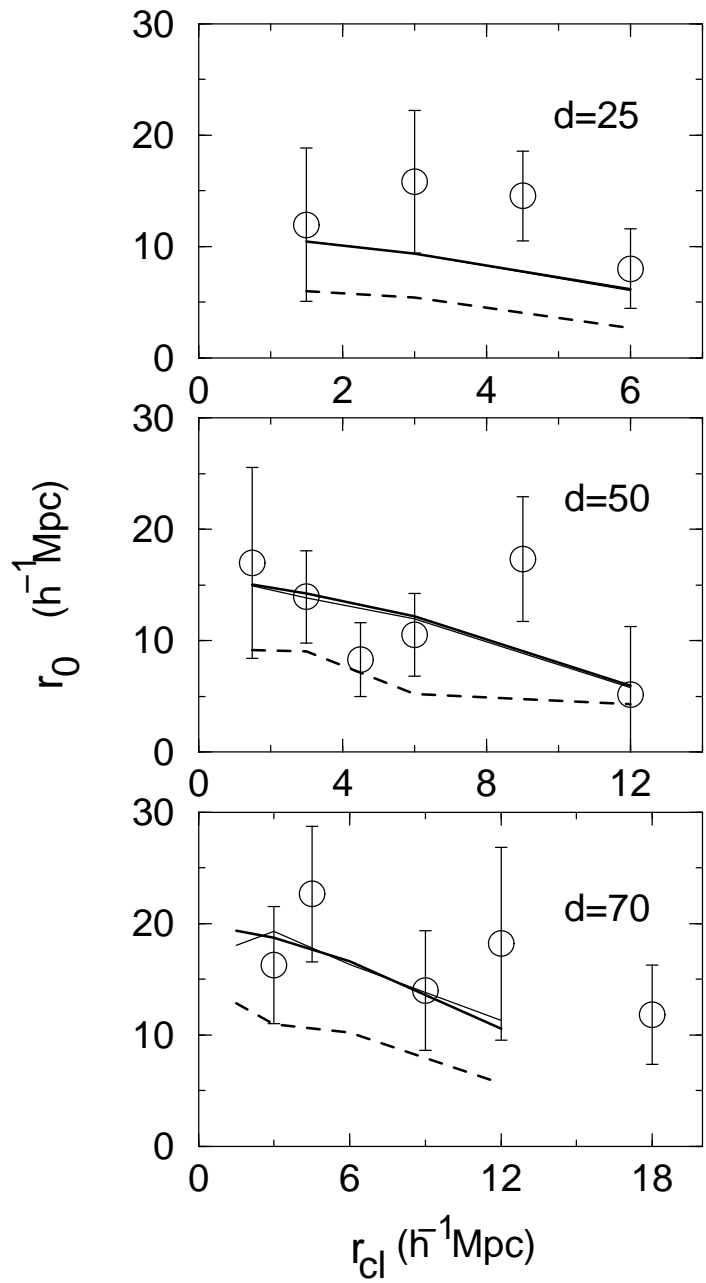
\end{document}